\documentclass[aps,prl,twocolumn,showpacs,superscriptaddress]{revtex4-1}  
\usepackage{graphicx}  
\usepackage{hyperref}
\usepackage{verbatim}

\usepackage{graphicx}  
\usepackage{dcolumn}   
\usepackage{bm}        
\usepackage{amssymb}   
\usepackage{amsmath}
\usepackage{lipsum}
\usepackage{xcolor}

\def\be{\begin{equation}}
\def\ee{\end{equation}}

\begin{document}
\def\brho{{\boldsymbol \rho}}
\def\bchi{{\boldsymbol \chi}}
\def\r{{\bf r}}
\def\ba{{\bf a}}
\def\bb{{\bf b}}
\def\bc{{\bf c}}
\def\bw{{\bf w}}
\def\bd{{\bf d}}
\def\bk{{\bf k}}
\def\bp{{\bf p}}
\def\bq{{\bf q}}
\def\br{{\bf r}}
\def\bv{{\bf v}}
\def\bx{{\bf x}}
\def\bz{{\bf z}}
\def\bG{{\bf G}}
\def\bR{{\bf R }}
\def\bH{{\bf H}}
\def\bF{{\bf F}}
\def\bJ{{\bf J}}
\def\bP{{\bf P}}
\def\la{\langle}
\def\ra{\rangle}
\def\calH{\mathcal{H}}
\def\calM{\mathcal{M}}
\def\calG{\mathcal{G}}
\def\calL{\mathcal{L}}
\def\calV{\mathcal{V}}
\def\calE{\mathcal{E}}
\def\calZ{\mathcal{Z}}
\def\p{\hat {\psi}}
\def\pd{\hat {\psi}^{\dag}}
\def\grad{\mbox{\boldmath $\nabla$}}
\def\Tr{{\rm Tr}}
\def\e{\epsilon}
\def\ve{\varepsilon}
\def\pa{\partial}
\def\nn{\nonumber}
\def\t{\tau}
\def\kbar{\bar {k}}
\def\rbar{\bar {r}}
\def\nbar{\bar {n}}

\title{Long range mediated interactions in a mixed dimensional system}

\author{Daniel Suchet}
\affiliation{Laboratoire  Kastler  Brossel,  ENS-PSL  Research  University,CNRS,  UPMC,  Coll\`{e}ge  de  France,  24,  rue  Lhomond,  75005  Paris}

\author{Zhigang Wu}
\email{zwu@mail.tsinghua.edu.cn}
\affiliation{Institute for Advanced Study, Tsinghua University, Beijing, 100084, China}

\author{Fr\'{e}d\'{e}ric  Chevy}
\affiliation{Laboratoire  Kastler  Brossel,  ENS-PSL  Research  University,CNRS,  UPMC,  Coll\`{e}ge  de  France,  24,  rue  Lhomond,  75005  Paris}

\author{Georg M.~Bruun}
\affiliation{Department of Physics and Astronomy, Aarhus University, DK-8000 Aarhus C, Denmark}
\date{\today}

\begin{abstract}
We present a mixed-dimensional atomic gas system to unambiguously detect and systematically probe mediated interactions. In our scheme, fermionic atoms are confined in two parallel planes and interact via exchange of elementary excitations in a three-dimensional background  gas. This interaction gives rise to a frequency shift of the out-of-phase dipole oscillations of the two clouds, which we calculate using a strong coupling theory taking the two-body  mixed-dimensional scattering into account exactly. The shift is shown to be easily measurable for strong interactions and can be used as a probe for mediated interactions.

 \end{abstract}

\maketitle
Mediated interactions  were originally introduced to provide a quantum-mechanical explanation for the peculiar ``action at a distance"  interactions like gravity and electromagnetism and they now constitute a major overarching paradigm in physics. In particle physics, exchange of gauge bosons is responsible for the propagation of fundamental interactions~\cite{weinberg1995}. In condensed matter, the attraction between the electrons in BCS superconductors arises from the exchange of lattice phonons~\cite{schrieffer1983}, and it is speculated that the mechanism
behind high-$T_c$ superconductivity lies in the exchange of spin fluctuations~\cite{Scalapino1995}. The concept of mediated interactions is also important in classical physics, where fluctuations of classical fields are responsible for phenomena such as the finite-temperature Casimir effect in electrodynamics \cite{milton2001casimir} and in  biophysics \cite{machta2012critical}.

Ultracold atoms have emerged as a versatile platform for the  investigation of many-body physics, and a host of schemes have been proposed to explore mediated interactions using these systems. For instance, mediated interactions lead to the formation of a $p$-wave superfluid  in spin-imbalanced fermionic systems \cite{bulgac2006ipw,lobo2006nsp,Mora2010Normal,yu2010comment}; they are responsible for the formation of a topological superfluid with a high critical temperature in 2D systems~\cite{Wu2016,Midtgaard2016,Caracanhas2017}, and in 1D quantum liquids they are shown to result in Casimir-like forces between impurities~\cite{schecter2014phonon}. In most cases, however, the mediated interaction is weak
and in competition with direct interactions between atoms, making its experimental observation challenging.

\begin{figure}
\includegraphics[width=\columnwidth]{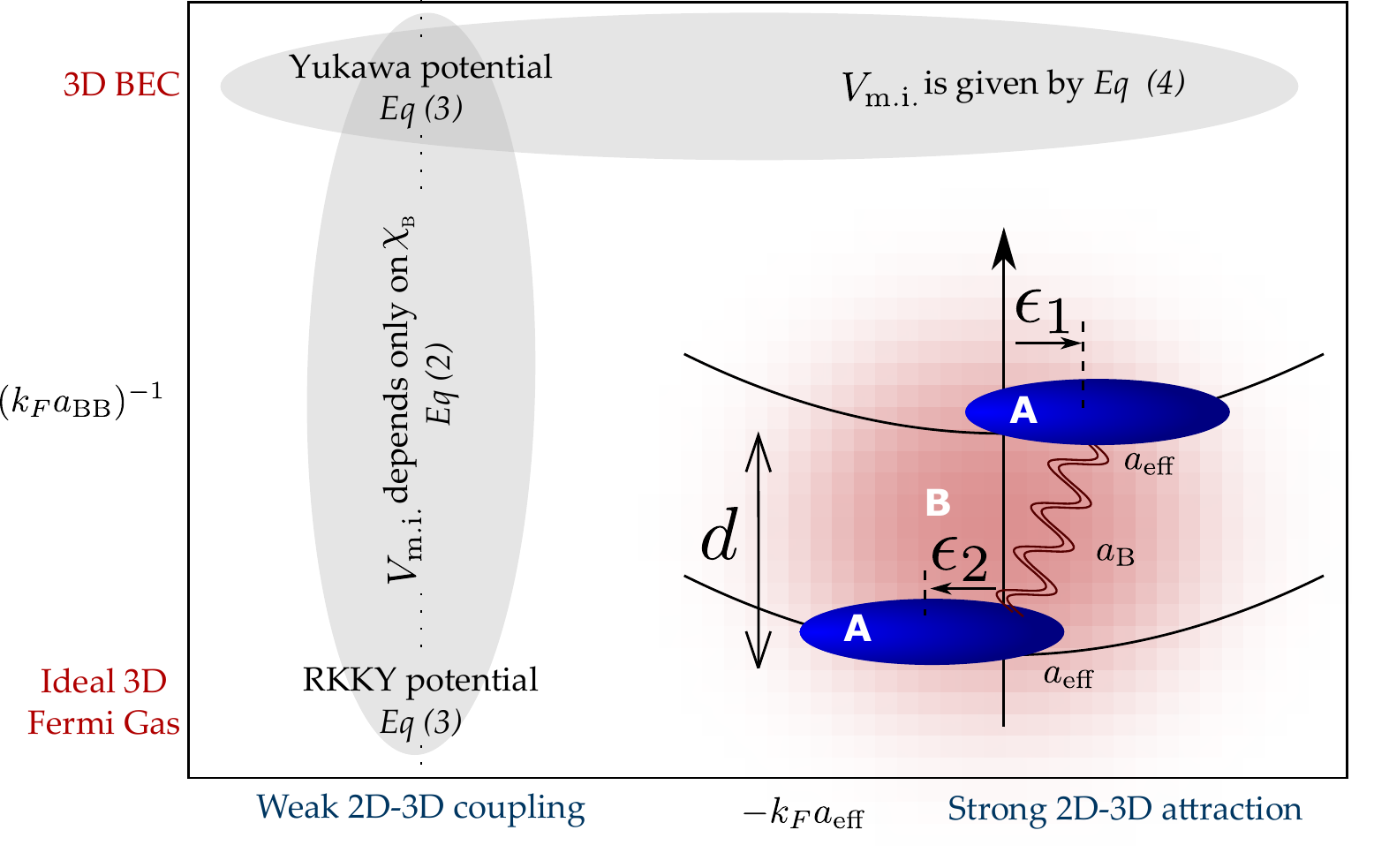}
\caption{We consider A-fermions confined in two layers by two identical  harmonic traps with a frequency $\omega_z$ much larger than any other energy scale in the system, and trapped in the $xy$ plane by a weak harmonic potential  with frequency $\omega_\perp$. The two layers are immersed in a 3D cloud of spin $1/2$ fermions (B-atoms), which
mediates an interaction between  the two layers. This mediated interaction gives rise to a frequency shift of the out-of-phase dipole
oscillation of the two A-clouds, which depends on the  B-B  scattering length $a_{BB}$ as well as on the  2D-3D A-B scattering length $ a_{\rm eff}$.
The ranges of $a_{\rm eff}$ and $a_{BB}$ analyzed in this paper are indicated by the grey regions. The main focus of our paper is for a strong 2D-3D interaction and on the BEC side of the 3D gas with a dimer scattering length $a_B$. 
 }

\label{fig:setup}
\end{figure}

In this paper, we apply the mixed-dimensional setup proposed in \cite{Nishida2010} and illustrated in Fig. \ref{fig:setup} to study mediated interactions.  Specifically we consider two parallel layers located at $z_1=0$ and $z_2=d$, which contain an equal number of spin-polarized non-interacting fermions (A-species). The layers are immersed in a uniform 3D gas of interacting spin $1/2$ fermions (B species), which can be tuned through the BEC-BCS cross-over. The presence of the 3D gas induces a mediated interaction between the A-particles: one A-particle will perturb locally the surrounding B-particles thereby inducing excitations in the 3D gas, which in turn affects the dynamics of a second A-particle. If  A-particles are harmonically trapped, this mediated coupling leads to a beating between oscillations in the two planes. Measuring the beating frequency between the 2D-clouds therefore gives access to the strength of mediated interaction. This scheme is similar to Coulomb drag experiments in bilayered electronic systems \cite{rojo1999electron} that was recently generalized to the case of dipolar gases \cite{matveeva2011dipolar}.

To analyze the dynamics of this system, we develop a systematic many-body theory for the  mediated inter-plane interaction that includes the low-energy mixed-dimensional A-B scattering exactly.  We then derive an  expression for the associated interaction energy between the two planes and calculate
 the frequency of the out-of-phase dipole oscillations of the 2D clouds in the $xy$-plane. In the weak A-B interaction limit, our results recover the perturbative expression for a mediated interaction proportional to the  density-density response function of the 3D gas.
 In the strong A-B interaction limit, however, the weak-coupling result breaks down completely. In the latter case we focus on the BEC regime of the 3D gas and show that the mediated interaction gives rise to a significant and easily detectable shift in the out-of-phase dipole oscillation frequency of the two clouds.

\emph{2D-3D scattering.--}
The interaction between the A and B particles is short range and
can  be characterised by an effective 2D-3D scattering length $a_{\rm eff}$~\cite{nishida2008universal}. Solving for
the scattering matrix in the many-body medium yields
\begin{align}
\mathcal T_{AB}(\bp_\perp,i\omega_\nu) = \frac {g}{1- g\Pi(\bp_\perp,i\omega_\nu)},
\label{TAB}
\end{align}
where  $g=2\pi a_\text{eff}/\sqrt{m_Bm_r}$, and $m_r=m_Am_B/(m_A+m_B)$ is the reduced mass ($\hbar=k_B=1$). Here $m_A$ denotes the mass of an A-fermion and $m_B$ that of the scattering particle in the 3D gas, namely the mass of B-fermion (dimer) in the BCS (BEC) regime. $\Pi(\bp_\perp,i\omega_\nu)$ is the renormalised 2D-3D pair propagator for the center-of-mass (COM) momentum $\bp_\perp=(p_x,p_y)$ in the plane,  and  $i\omega_\nu$ is either a bosonic (BCS regime) or fermionic (BEC regime) Matsubara frequency.  Equation (\ref{TAB}) includes many-body effects in the ladder approximation (see the Supplemental Material), and recovers the correct low energy  2D-3D scattering matrix in a vacuum~\cite{Nishida2009}. .

\emph{Mediated interaction for weak 2D-3D interaction.--}
Consider first the case of a weak 2D-3D interaction where $a_\text{eff}$ is much smaller than the interparticle spacing of the A and B particles.
We  then have $T_{AB}(\bp_\perp,i\omega_\nu)\simeq g$ from  (\ref{TAB}), and second-order perturbation theory gives
\begin{align}
 V_{\rm m.i.}({\mathbf q_\perp},i\omega_\nu)=g^2\int_{-\infty}^\infty \!d q_z e^{iq_zd} \chi_B\left(\bq_\perp,q_z,i\omega_\nu\right),
 \label{Eq:Veff0}
 \end{align}
which describes the mediated interaction between two A-particles in  different planes. Here $(\bq_\perp,i\omega_\nu)=(q_x,q_y,i\omega_\nu)$
 are the transferred momentum and frequency and  $\chi_B(\bq_\perp,q_z,i\omega_\nu)$ is the density-density response function of the  B-cloud.
 The integration over the momentum $q_z$  comes from the fact that it is not conserved in the 2D-3D scattering. Deep in the BCS  limit where the
B fermions form an ideal Fermi gas,  the mediated interaction (\ref{Eq:Veff0}) is of the form of a Ruderman-Kittel-Kasuya-Yosida potential \cite{RKKY1,RKKY2,RKKY3,Nishida2010}.
When the B fermions are deep in the BEC limit where they form a weakly interacting BEC of dimers, the mediated interaction takes the form of a Yukawa  potential \cite{Yukawa}. At zero frequency, Fourier transforming (\ref{Eq:Veff0}) back to the real space gives
\be
V_{\rm m.i.}(r) =
\begin{cases}
  g^{2}\frac{m_B }{16\pi^{3}}\frac{2p_Fr\cos2p_Fr-\sin2p_Fr}{r^{4}}&\text{BCS limit}
  \\
  -g^{2}\frac{n_Bm_B }{\pi r}e^{-\sqrt{2}r/\xi_B}  &\text{BEC limit}
  \label{eq:Yukawa}
  \end{cases}
\ee
 where $p_F$ is the Fermi momentum of the 3D Fermi gas in the BCS regime, and  $n_B$ is the density of the 3D BEC of dimers with coherence length $\xi_B=1/\sqrt{8\pi n_B  a_B}$.
 Here, $ a_B=0.6a_{BB}$ is the  scattering length between the  deeply bound dimers of B fermions~\cite{Petrov2004}.

\emph{Mediated interaction for strong 2D-3D interaction.--}
For a strong 2D-3D interaction where $a_\text{eff}$ is comparable to or larger than the interparticle spacing, the mediated interaction between the two layers  takes on a more complex form. The reason is that we need to retain the  full COM momentum and frequency dependence of the 2D-3D scattering matrix given by (\ref{TAB}).

\begin{figure}[ht]
	\centering
	
	\includegraphics[width=1\columnwidth]{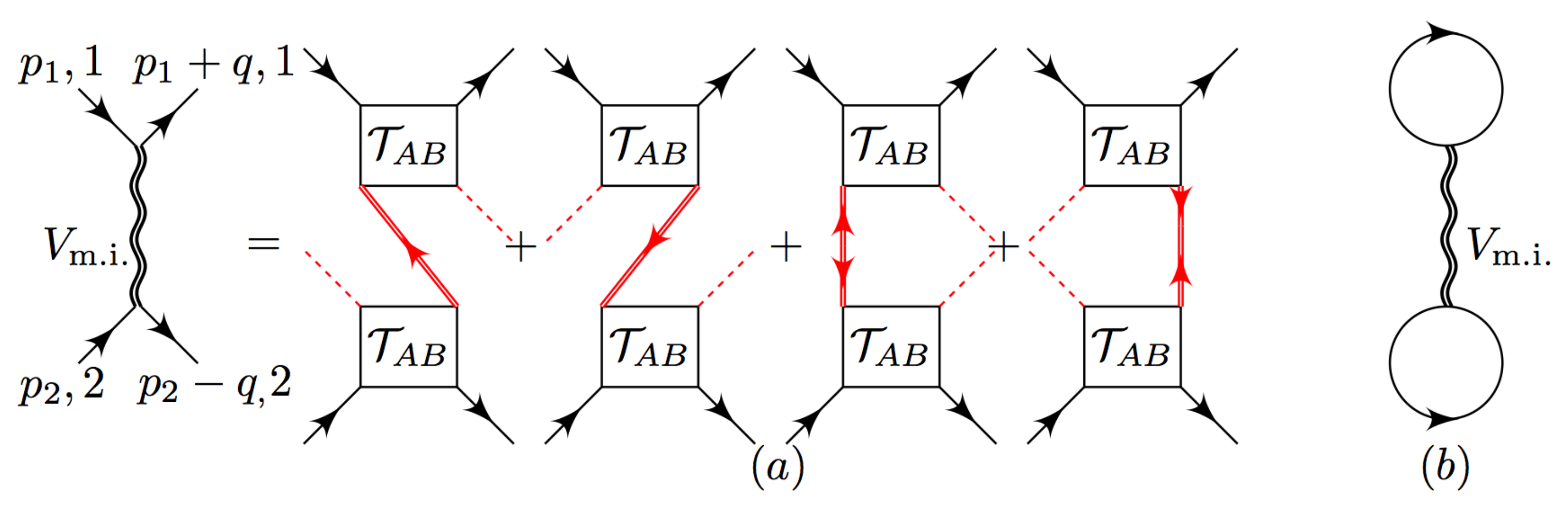}
	\caption{(a) The mediated interaction $V_{\rm m.i.}$ between  fermions in layer $1$ and $2$ coming from the exchange of one Bogoliubov mode. The
	 box represents the 2D-3D scattering amplitude $\mathcal {T}_{AB}$, the dashed red line represents bosons emitted or absorbed by the condensate, and
	 the red thick line with one arrow
	 and two arrows represents the normal $G^\text{B}_{11}$ and anomalous Green's functions $G^\text{B}_{12}$ (or $G^\text{B}_{21}$) respectively.
	 (b) The leading correction to the thermodynamic potential due to the mediated interaction between the two planes. The thin solid lines represent the Fermi propagators in the two planes.}
	\label{fig:Vij}
\end{figure}

 We shall from now on concentrate on the BEC limit of the B-fermions, namely when they form a weakly interacting BEC of dimers, which can be treated within Bogoliubov theory.
 The mediated interaction between the A-particles is calculated including
 all processes where a single Bogoliubov phonon in the BEC is exchanged between  the two layers.
  In a diagrammatic language, these processes are  shown in Fig.~\ref{fig:Vij} (a).
Summing up the contributions from the four terms  in Fig.~\ref{fig:Vij} (a) gives
\begin{align}
V_{\rm m.i.}(p_{1},p_{2};q)&  = n_B\mathcal T_{AB}(p_{1}+q)\mathcal T_{AB}(p_{2}) \bar G^\text{B}_{11}(\mathbf q_\perp,i\omega_\nu)\nn \\
+&n_B\mathcal T_{AB}(p_{1})\mathcal T_{AB}(p_{2}-q) \bar G^\text{B}_{11}(-\mathbf q_\perp,-i\omega_\nu)\nn \\
+&n_B\mathcal T_{AB}(p_{1}+q)\mathcal T_{AB}(p_{2}-q) \bar G^\text{B}_{12}(\mathbf q_\perp,i\omega_\nu)\nn \\
+&n_B\mathcal T_{AB}(p_{1})\mathcal T_{AB}(p_{2}) \bar G^\text{B}_{21}(\mathbf q_\perp,i\omega_\nu),
\label{ind_interaction_1}
\end{align}
where  $p_{1} \equiv (\bp_{1\perp},i\omega_{m_1})$, $p_{2} \equiv (\bp_{2\perp},i\omega_{m_2})$, and $q \equiv (\bq_\perp,i\omega_{\nu})$.
Here $\omega_m = (2m+1)\pi /\beta$ and $\omega_\nu = 2\nu\pi /\beta$ are Fermi and Bose Matsubara frequencies respectively, where $\beta=1/T$ is the inverse temperature and  $m$ and $\nu$ are integers.
In  (\ref{ind_interaction_1}), the Green's functions of the BEC are integrated over the $z$-component of the momentum as
\begin{align}
\bar G^\text{B}_{\alpha\beta}(\mathbf q_\perp,i\omega_\nu)\equiv \int_{-\infty}^\infty\frac{dq_z}{2\pi}G^\text{B}_{\alpha\beta}(\mathbf q_\perp,q_z,i\omega_\nu)e^{iq_z d}.
\label{kzFourier}
\end{align}
The Green's functions of the 3D BEC are as usual
\be
G^\text{B}_{11} = \frac{u_{\bk}^2}{i\omega_\nu-E_{\bk}}-\frac{v_{\bk}^2}{i\omega_\nu+E_{\bk}}, \hspace{0.3cm}G^\text{B}_{12}=\frac{g_Bn_B}{\omega_\nu^2+E_{\bk}^2}
\label{GB11}
\ee
where $\bk = (\bk_\perp,k_z)$ and $G^\text{B}_{21}(\bk,i\omega_\nu)=G^\text{B}_{12}(\bk,i\omega_\nu)$. We have defined
   $u_{\bk}^2, v_{\bk}^2 ={\frac{1}{2} \left[ \left ({\e_{\bk} + g_Bn_B}\right)/{E_{\bk}} \pm 1 \right]}$, $E_\bk = \sqrt{ \e_{\bk} (\e_{\bk} + 2 g_Bn_B) }$ is the Bogoliubov spectrum with
  $\e_\bk = k^2/2m_B$, and  $g_B = 4\pi a_B/m_B$. Note that the mediated interaction
   (\ref{ind_interaction_1}) depends on both  $p_1$ and $p_2$ as well as $q$ due to the momentum and frequency dependence of the
 2D-3D scattering. In fact, in the weak interaction limit $\mathcal{T}_{AB}\simeq g$, one recovers (\ref{Eq:Veff0}) from the more general expression (\ref{ind_interaction_1}).

\emph{Thermodynamical potential.--}
We now derive an expression for the correction to the thermodynamic potential $\Omega$ due to the mediated interaction
between the two planes for a general strength of the  2D-3D interaction. The dominant contribution is the Hartree term illustrated in Fig.~\ref{fig:Vij} (b). For a
homogeneous system, this term gives the correction per unit area as(for the rest of the paper the $\perp$ subscript will be dropped in the vector notation and all bold face letters now denote in-plane 2D vectors)
\begin{gather}
 \bar \Omega_{\rm m.i.} = \frac{1}{\beta^2}\sum_{m_1 m_2}\int \frac{d^2p_{1}}{(2\pi)^2}\frac{d^2p_{2}}{(2\pi)^2}V_\text{\rm m.i.}(p_1,p_2;0) \nn \\
 \times G_{1}^\text{A}(\bp_{1},i\omega_{m_1})G^\text{A}_{2}(\bp_{2},i\omega_{m_2}),
\label{deltaomega}
\end{gather}
where 
$
G^\text{A}_{j}(\bp,i\omega_m) = {1}/({i\omega_m-p^2/2m_A +\mu_A})
$
is the Green's function for the A-fermions in the $j$-th layer with  $\mu_A$ being the chemical potential. Using (\ref{ind_interaction_1})
together with the identity $2\bar G_{11}(0,0)+2\bar G_{12}(0,0)=-\sqrt2n_Bm_B\xi_B\exp(-\sqrt2d/\xi_B)$ yields
\begin{align}
 \bar \Omega_{\rm m.i.} =-\sqrt{2}m_B\xi_B n_B e^{-\sqrt{2}d/\xi_B} \bar \Omega_1 \bar \Omega_2,
 \label{Omegamiuniform}
\end{align}
where
\begin{align}
 \bar \Omega_{j} = \frac{1}{\beta}\sum_{m}\int \frac{d^2p}{(2\pi)^2}\mathcal T_{AB}(\bp,i\omega_m)G^\text{A}_{j}(\bp,i\omega_m).
\label{Omegaa}
\end{align}
We point out that the Matsubara frequency summation in the above expression can in fact be performed analytically (see supplementary material), which greatly simplifies the numerical calculation of thermodynamic potential density. 

\emph{Local-density approximation.--} Using the local-density approximation, we can
 generalize (\ref{deltaomega}),
 which was derived assuming homogeneous system, to the case of trapped 2D Fermi clouds. This yields the total correction as 
  \begin{align}
\Omega_{\rm m.i.}(\e_1-\e_2) =\int d^2r_{1}d^2r_{2}[2\bar G^\text{B}_{11}(\br_{1}-\br_{2},0) \nonumber\\
+2\bar G^\text{B}_{12}(\br_{1}-\br_{2},0)] \bar \Omega_1 (\br_{1}-\e_1\hat \bx)  \bar \Omega_2 (\br_{2}-\e_2\hat \bx),
\label{omegadldaf}
\end{align}
where  $\bar G^\text{B}_{ij}(\br,0)$ is the Fourier transform of $\bar G^\text{B}_{ij}(\bp,0)$ back to real 2D space,
and $\bar \Omega_i (\br)$ is given by (\ref{Omegaa}) using a local chemical potential $\mu_A(\br)=\mu_A+m_A\omega_\perp^2r^2/2$.
In (\ref{omegadldaf}), we have allowed the two A-clouds to be rigidly displaced distances of $\e_1$ and $\e_2$ along the $x$-axis in order to analyse their coupled dipole
oscillations, see Fig.\ \ref{fig:setup}.  Since $\bar G^\text{B}_{ij}$ already contains a Fourier transform with respect to $z$-momentum, see (\ref{kzFourier}),
the bosonic Green's functions entering  (\ref{omegadldaf})  now simply add up  to the density-density correlation function of the BEC
evaluated at the 3D real space distance $r=|\br_{1}-\br_{2}+d\hat \bz|$. Using this, we finally obtain
\begin{align}
\Omega_{\rm m.i.}(\e_1-\e_2)=&-\frac{m_Bn_B}{\pi }\int d^2r_{1}d^2r_{2}  \frac{e^{-\sqrt2r/\xi_B}}r  \nn \\
&\times\bar \Omega_1 (\br_{1}-\e_1\hat \bx)  \bar \Omega_2 (\br_{2}-\e_2\hat \bx).
\label{omegadldafFinal}
\end{align}
Equation (\ref{omegadldafFinal}) can be understood as follows. Consider two area elements of the 2D gases,  one located at $\br_{1}-\e_1\hat \bx$ in  layer $1$ and the other at $\br_{2}-\e_2\hat \bx$ in layer $2$. The contribution from these two elements can be approximated by the expression in (\ref{Omegamiuniform}) in which the relative distance is taken to be $r$ instead of $d$. Equation (\ref{omegadldafFinal}) then sums up all such contributions in the two clouds.

For weak interaction, we see from (\ref{Omegaa}) that $\bar \Omega_{j}(\br_j-\e_j\hat \bx)=gn_j(\br_{j}-\e_j\hat \bx)$, where $n_j (\br_{j}-\e_j\hat \bx)$ denotes the equilibrium fermion density in layer $j$ rigidly displaced the distance $\e_j$ along the $x$-axis. Equation  (\ref{omegadldafFinal}) then simplifies to
\begin{align}
\Omega_{\rm m.i.}(\e_1-\e_2)=&-g^2\frac{m_Bn_B}{\pi }\int d^2r_{1}d^2r_{2}  \frac{e^{-\sqrt2r/\xi_B}}r  \nn \\
&\times n_1 (\br_{1}-\e_1\hat \bx) n_2(\br_{2}-\e_2\hat \bx),
\label{dEweak}
\end{align}
 which  is the usual Hartree approximation for the interaction energy between the two planes mediated by a Yukawa interaction.

\emph{Coupled dipole oscillations.--}
Consider now the situation where the two clouds perform dipole oscillations around their equilibrium positions, see Fig.~\ref{fig:setup}.
For small displacements $\e_1$ and $\e_2$, the COM velocities and the beating frequencies are  small compared to the speed of sound in the 3D gas and the trapping frequencies respectively, yielding rigid and undamped oscillations of the 2D clouds~\cite{Ferrier2014Mixture}. The COM dynamics is then determined by the energy increase $\delta E$  associated with the  displacements of the clouds.
For rigid displacements, we have
$\delta E = \Omega_{\rm m.i.}(\e_1-\e_2)-\Omega_{\rm m.i.}(0)+ [\mu_A(\e_1)+\mu_A(\e_2)-2\mu_A]N_A$, which gives
\begin{equation}
  \delta E(\e_1,\e_2)=  \frac{1}{2}N_Am_A\omega_\perp^2(\e_1^2 +  \e_2^2)  + \Omega_{\rm m.i.}(\e_1-\e_2)-\Omega_{\rm m.i.}(0),
\label{deltaE}
\end{equation}
where  $N_A$ is the number of fermions in each layer.
Taylor expanding $\Omega_{\rm m.i.}(\e_1-\e_2)$ to second order in $\e_1-\e_2$, we readily see that the motion of the two clouds separates into an in-phase oscillation with
frequency $\omega_\perp$, and an out-of-phase oscillation with frequency
\begin{align}
\omega_r = \omega_\perp \sqrt{1+ {2I}/{N_Am_A\omega_\perp^2}},
\label{omega_r}
\end{align}
where
\be
I = \left.\frac{\pa^2}{\pa \e_1^2}  \Omega_{\rm m.i.}(\e_1-\e_2)\right |_{\e_1-\e_2=0}.
\label{I}
\ee
The microscopic expression for $\omega_r $ for \emph{arbitrary} strength of the  2D-3D interaction in terms of (\ref{Omegaa}), (\ref{omegadldafFinal}), (\ref{omega_r}), and (\ref{I}) is the  main result of this letter and it explicitly shows how the mediated interaction can be probed by measuring the frequency of the out-of-phase
dipole oscillations of the two clouds.

\begin{figure}[ht]
	\centering
        \includegraphics[width=0.99 \columnwidth]{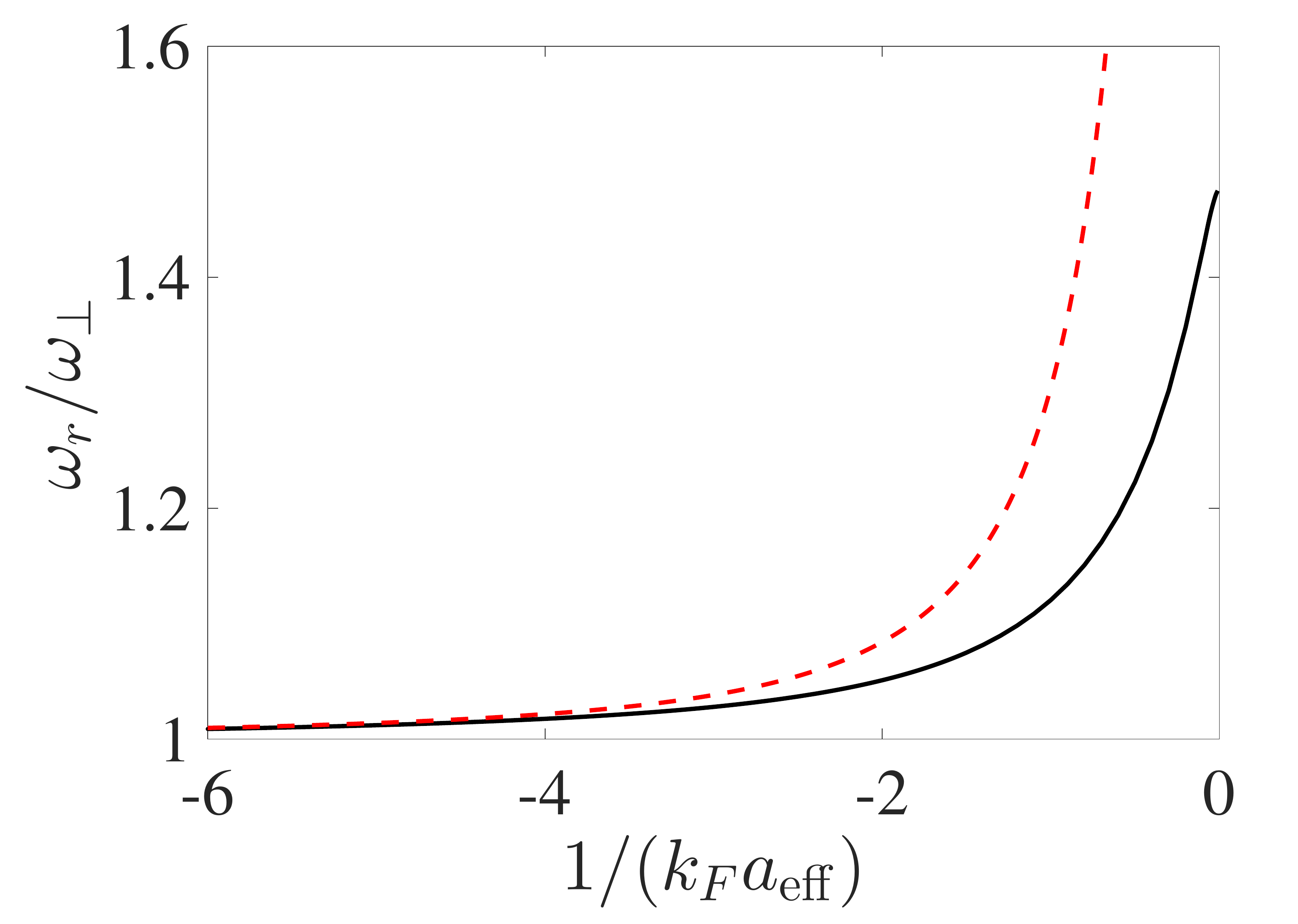}
	\caption{The frequency ratio $\omega_r/\omega_\perp$ of the out-of-phase dipole oscillation  as a function of $1/(k_Fa_{\rm eff})$. The solid line is the full strong coupling result whereas the dashed line is determined by the second order perturbation theory.}
	\label{fig:omegaraeff}
\end{figure}

\emph{Results.--}
We now calculate the frequency  $\omega_r$ for a realistic cold-atom system consisting of  $N_A = 1000$
 $^{40} \rm K$ atoms  trapped in each plane, immersed in a 3D BEC of $^6{\rm Li}$ dimers.
  The  transverse trapping frequency
 for the  $^{40} \rm K$ clouds is $\omega_\perp = 2\pi\times 380 \,{\rm Hz}$, the density of the BEC is $n_B=10^{18} \, m^{-3}$, and the coherence length is  $\xi_B= 2.7 \, \mu {\rm m} $. We furthermore assume that the temperature is zero. In Fig.~\ref{fig:omegaraeff}, we show the frequency  $\omega_r/\omega_\perp$ as a function of the 2D-3D interaction strength  $1/k_Fa_{\rm eff}$ at a fixed interlayer distance $d = 0.4 \, \mu {\rm m}$.
  The frequency increases monotonically as $a_{\rm eff}$ increases. For weak interaction, it agrees with the second order result (dashed line). For stronger interaction, the full frequency/momentum dependence of the 2D-3D scattering is important, and the perturbative result deviates significantly from the full strong-coupling theory. In particular, whereas the perturbative result
 diverges for $1/k_Fa_{\rm eff}\rightarrow 0$, the strong-coupling theory predicts a finite frequency saturating at $\omega_r\simeq 1.48 \, \omega_\perp$. Importantly, the frequency  shift
  becomes significant for $-2\lesssim1/k_Fa_{\rm eff}\le 0$, which includes a region sufficiently far from unitarity so that
   the predicted 3-body loss is small~ \cite{nishida2011liberating}. This demonstrates   the usefulness of our proposal to detect  mediated interactions.
   Note that this result can only be obtained using a strong coupling theory, since the perturbative result is only accurate for weak interactions where the frequency shift is minute.
\begin{figure}[ht]
	\centering
        \includegraphics[width=0.99 \columnwidth]{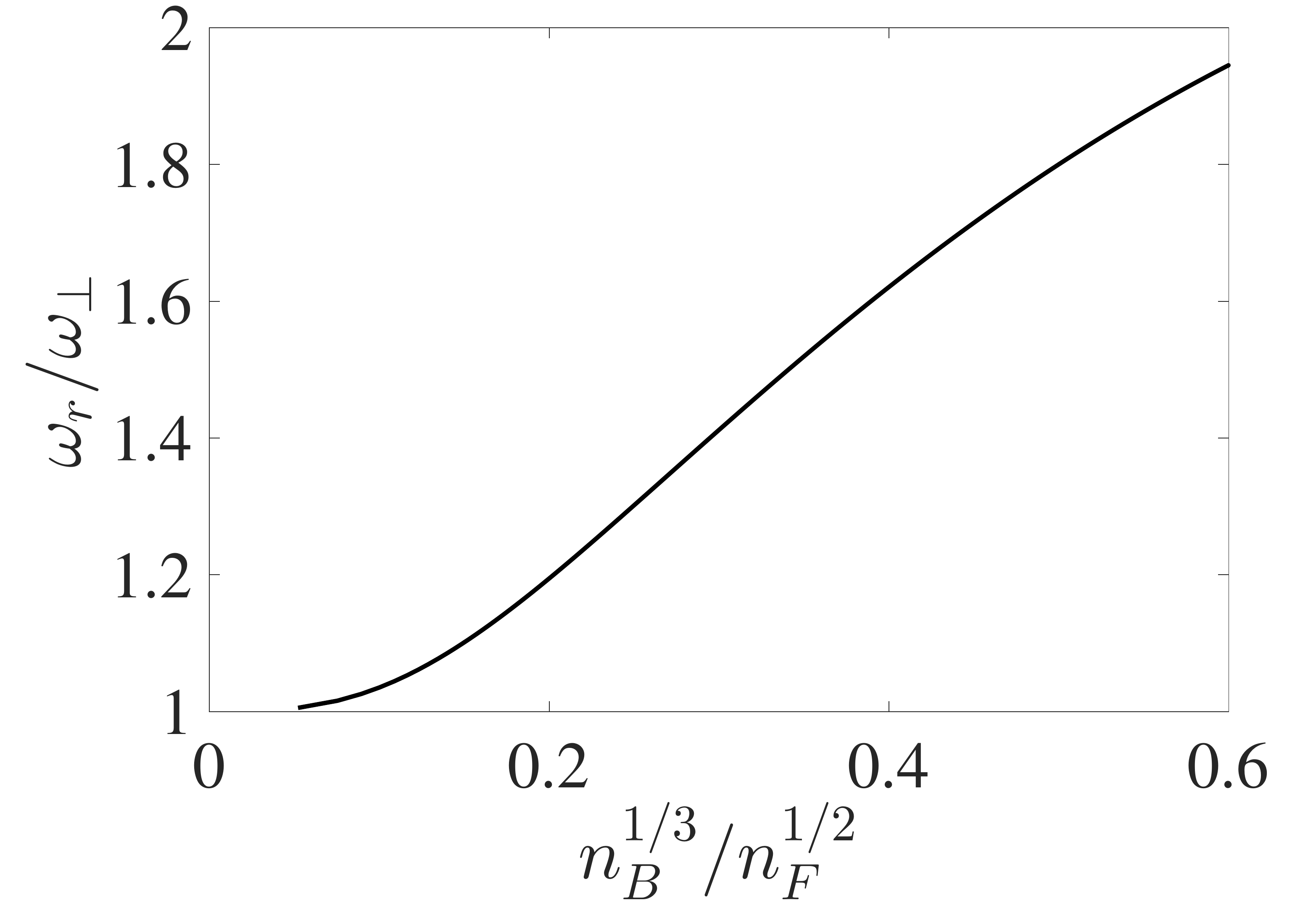}
	\caption{The frequency ratio $\omega_r/\omega_\perp$  as a function of the ratio of the interparticle distances $n_B^{1/3}/n_F^{1/2}$ (keeping $n_F$ fixed) for $1/(k_Fa_{\rm eff})=-0.1$	and all other parameters as in Fig.\ 3. Here $n_F$ is the fermion density at the center of the cloud.}
	\label{fig:omegaraeff2}
\end{figure}

In Fig.~\ref{fig:omegaraeff2}, we plot $\omega_r/\omega_\perp$ as a function of the ratio of the interparticle distances $n_B^{1/3}/n_F^{1/2}$ (keeping $n_F$ fixed)  with $1/k_Fa_{\rm eff}=-0.1$ and all other physical  parameters the same as for  Fig.~\ref{fig:omegaraeff}.a. The density of the BEC enters the mediated interaction in two ways, which is most clearly seen in the weak-coupling limit given by  (\ref{eq:Yukawa}): First, the strength of the  interaction is  proportional to $n_B$;  second, the range of the  interaction is determined by the BEC coherence length $\xi_B\propto 1/\sqrt{n_B}$. Thus, increasing the density increases the strength but reduces the range of the mediated interaction, and it is not a priori obvious what the net effect on the frequency shift will be. From Fig.~\ref{fig:omegaraeff2}, we see that for the chosen parameters, $\omega_r$ in fact increases monotonically with increasing BEC density \footnote{We restricted all figures to negative values of the 2D-3D scattering length. Indeed for $1/k_Fa_{\rm eff}> 0$, a 2D fermion can form a bound-dimer state with a 3D boson.
The frequency shift in this region therefore depends on whether the system forms these dimers, or whether it is  on the so-called repulsive branch where the effective 2D-3D interaction is repulsive. This complicates the analysis, which will be presented in a future publication.
}.

 \emph{Conclusions.--}
 We demonstrated that a mixed-dimensional setup consisting of two layers of identical fermions immersed in a 3D background gas is a powerful probe to investigate mediated interactions systematically. The mediated interaction between the two layers modifies the out-of-phase dipole oscillation frequency of the 2D clouds, and we calculate this shift using a strong-coupling theory taking into account the low energy scattering between the 2D and 3D particles. Using this theory, we showed that for strong 2D-3D coupling, the resulting frequency shift is clearly measurable.

Finally we note that the advantages of our proposal are twofold. First, if the 2D trapping is realized using optical potentials, the distance between planes is a few hundred nanometres, which is much larger than the range of interatomic interactions. Any observed coupling between the two planes is therefore solely due to a mediated interaction via the 3D gas. Second, the shift of the center-of-mass oscillation frequency is a very precise spectroscopic tool that can be used as a probe of  weak interactions, as demonstrated recently in \cite{Ferrier2014Mixture,roy2016two}.

\begin{acknowledgements}
FC and DS acknowledge support from R\'egion Ile de France (DIM IFRAF/NanoK), ANR (Grant SpiFBox) and European Union (ERC Grant ThermoDynaMix).
GMB and ZW wishes to acknowledge the support of the Villum Foundation via Grant No. VKR023163.

DS  and ZW contributed equally to this work.

\end{acknowledgements}

\widetext
\clearpage
\begin{center}
\textbf{\large Supplemental Material }\\
\vspace{4mm}
{Daniel Suchet$^1$, Zhigang Wu$^2$, Fr\'{e}d\'{e}ric  Chevy$^1$, and G. M. Bruun$^3$}\\
\vspace{2mm}
{\em \small
$^1$Laboratoire  Kastler  Brossel,  ENS-PSL  Research  University,CNRS,  UPMC,  Coll\`{e}ge  de  France,  24,  rue  Lhomond,  75005  Paris \\
$^2$Institute for Advanced Study, Tsinghua University, Beijing, 100084, China\\
$^3$Department of Physics and Astronomy, Aarhus University, DK-8000 Aarhus C, Denmark
}\end{center}

\section{2D-3D scattering matrix}
\begin{figure}[ht]
	\centering
	\includegraphics[width=0.8\textwidth]{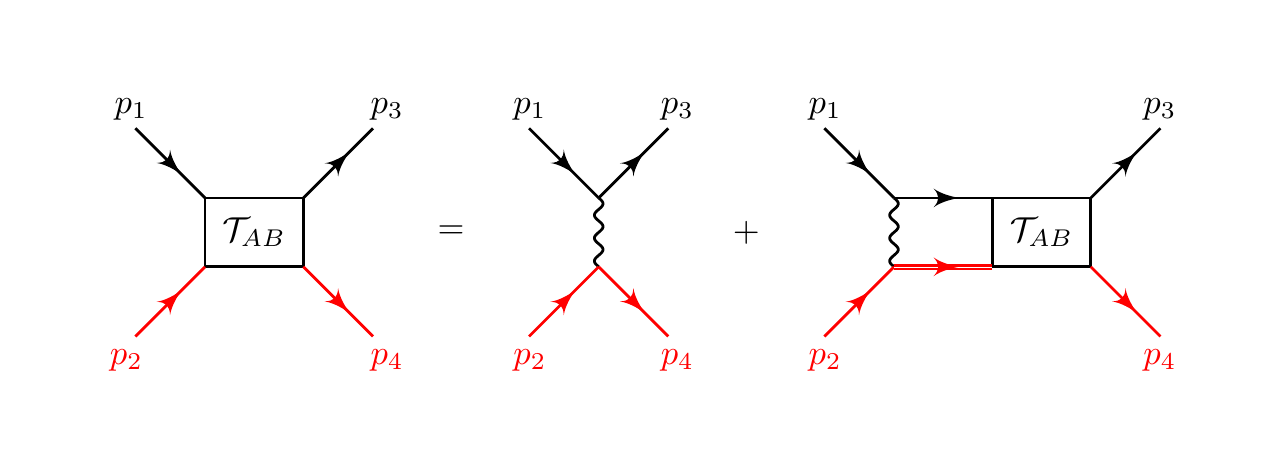}
	\caption{Scattering T-matrix between a 2D fermion and a 3D boson. The black line represents the femion propagator and the red thick line represents the normal boson propagator. Here $p_{1} \equiv (\bp_{1\perp},i\omega_{n_1})$, $p_{2} \equiv (\bp_{2},i\omega_{\nu_2})$, $p_{3} \equiv (\bp_{3\perp},i\omega_{n_3})$ and $p_{4} \equiv (\bp_{4},i\omega_{\nu_4})$.}
	\label{fig:tmatr}
\end{figure}

We provide some details on the 2D-3D scattering matrix given in Eq.~(1) in the main text.  In the strong 2D-3D interaction limit and in the presence of a 3D BEC background, we need the scattering amplitude in medium between the a 2D fermion and a 3D boson to determine the mediated interaction. In terms of the well-known T-matrix approximation, the scattering amplitude $\mathcal T_{AB}$ satisfies an integral equation represented diagrammatically in Fig.~\ref{fig:tmatr}. Here the $\perp$ subscript is used to distinguish 2D plane vectors from the 3D ones. Using standard procedure the scattering matrix can be expressed in terms of the 2D-3D zero-energy scattering amplitude in vacuum $ g=2\pi a_\text{eff}/{\sqrt{m_Bm_{r}}}$, where $m_r = m_Am_B/M$ with $M=m_A + m_B$ being the reduced mass and $a_{\rm eff}$ being the 2D-3D scattering length. In doing so, it can be shown that $\mathcal T_{AB}$ only depends on the total momentum and frequency $\bp_\perp=\bp_{1\perp}+\bp_{2\perp} = \bp_{3\perp}+\bp_{4\perp}$ and $\omega_\nu = \omega_{n_1} + \omega_{\nu_2} =  \omega_{n_3} + \omega_{\nu_4}$. We find
\begin{align}
\mathcal T_{AB}(\bp_\perp,i\omega_\nu) = \frac { g}{1- g\Pi(\bp_\perp,i\omega_\nu)}.
\label{TAB}
\end{align}
Here $\Pi(\bp_\perp,i\omega_\nu)$ is the renormalised pair propagation given by
\begin{align}
\Pi(\bp_\perp,i\omega_\nu) = \int \frac{d^3p'}{(2\pi)^3}  \left[
				u_{\bp_+}^{2}
				\frac{1 + b(E_{\bp_+}) - f(\xi_{\bp_-})}
					{i\omega_\nu - E_{\bp_+} - \xi_{\bp_-}}
				+
				v_{\bp_+}^{2}
				\frac{b(E_{\bp_+}) + f(\xi_{\bp_-})}
					{i\omega_\nu + E_{\bp_+} - \xi_{\bp_-}}+\frac {1}{p_z^{\prime 2}/2m_B+\bp'^2_\perp/2m_r+i0^+}\right ],
\end{align}
where $\bp'=(\bp'_\perp,p'_z)$, $\bp_+ \equiv \frac{m_B}{M}\bp_\perp+\bp'$, $\bp_- \equiv \frac{m_A}{M}\bp_\perp - \bp'_\perp$, and $b(x) = 1/(e^{\beta x}-1)$ and $f(x)=1/(e^{\beta x}+1)$ are the Bose and Fermi distribution function respectively.  For weakly interacting Bosons, it is a good approximation to replace the normal Green's function $G^{\rm B}_{11}(\bq,i\omega_\nu)$ by the non-interacting Boson Green's function $G^{\rm B}_{0}(\bq,i\omega_\nu) = 1/(i\omega_\nu-\e_{\bq}+\mu_B)$ in the scattering T-matrix. With this simplification, we find at $T=0$
\begin{align}
\Pi(\bp_\perp,i\omega_\nu) = \int \frac{d^3p'}{(2\pi)^3}  \left[
				\frac{1  - \theta\left (k_F-\left |m_A\bp_\perp/M-\bp'_\perp\right |\right )}
					{i\omega_\nu - \left ( \bp_\perp^2/2M+\bp_\perp^{\prime 2}/2m_r+p_z^{\prime 2}/2m_B\right )+\mu_A}+\frac {1}{p_z^{\prime 2}/2m_B+\bp^{\prime 2}_\perp/2m_r+i0^+}\right ],
\end{align}
where $k_F = \sqrt{2m_A\mu_A}$ is the Fermi momentum of the A-species. 
Expressed in terms of the  dimensionless variables, the pair propagator is 
\begin{align}
\Pi(\bp_\perp,i\omega_\nu) = 2m_A k_F\int \frac{d^2 p_\perp'}{(2\pi)^2}\int \frac{d p'_z}{2\pi}  \left[
				\frac{1  - \theta\left (1-\left |\bp'_\perp-\alpha_A\bp_\perp\right |\right )}
					{i\omega_\nu - \left ( \alpha_Ap_\perp^2+\alpha_B^{-1}p_\perp^{\prime 2}+\alpha_A\alpha_B^{-1}p_z^{\prime 2}\right )+1}+\frac {1}{\alpha_A\alpha_B^{-1}p_z^{\prime 2}+\alpha_B^{-1}p_\perp^{\prime 2}+i0^+}\right ],
\end{align}
where  $\alpha_A = m_A/M$ and $\alpha_B = m_B/M$.  Here the frequency variables are scaled in terms of the chemical potential $\mu_A$ and the momentum variables in terms of the Fermi momentum $k_F$. We write 
\be
\Pi(\bp_\perp,i\omega_\nu)  =  \Pi_0(\bp_\perp,i\omega_\nu) + \Delta \Pi(\bp_\perp,i\omega_\nu),
\label{Piadd}
\ee
where
\begin{align}
 \Pi_0(\bp_\perp,i\omega_\nu)  &\equiv 2m_A k_F\int \frac{d^2 p'_\perp}{(2\pi)^2}\int \frac{d p'_z}{2\pi}  \left[
				\frac{1 }
					{i\omega_\nu - \left ( \alpha_Ap_\perp^2+\alpha_B^{-1}p_\perp^{\prime 2}+\alpha_A\alpha_B^{-1}p_z^{\prime 2}\right )+1}+\frac {1}{\alpha_A\alpha_B^{-1}p_z^{\prime 2}+\alpha_B^{-1}p_\perp^{\prime 2}+i0^+}\right ] \nn \\
					& = -i\frac{m_A k_F}{2\pi{\alpha_A}^{1/2}\alpha_B^{-3/2}}\sqrt{i\omega_\nu+1-\alpha_Ap_\perp^2}
					\label{Pi0}
\end{align}
is the pair propagator in vacuum and 
\begin{align}
\Delta \Pi(\bp_\perp,i\omega_\nu) &\equiv -2m_A k_F\int \frac{d^2 p'_\perp}{(2\pi)^2}\int \frac{d p'_z}{2\pi}
				\frac{ \theta\left (1-\left |\bp'_\perp-\alpha_A\bp_\perp\right |\right )}
					{i\omega_\nu - \left ( \alpha_Ap_\perp^2+\alpha_B^{-1}p_\perp^{\prime 2}+\alpha_A\alpha_B^{-1}p_z^{\prime 2}\right )+1} \nn \\
					& = i\frac{m_A k_F}{\sqrt{\alpha_A\alpha_B^{-1}}}\int \frac{d^2 p'_\perp}{(2\pi)^2}\frac{\theta\left (1-\left |\bp'_\perp-\alpha_A\bp_\perp\right |\right )}{\sqrt{i\omega_\nu+1-\alpha_Ap_\perp^2-\alpha_B^{-1}p_\perp^{\prime 2}}}
					\label{DPi}
\end{align}
is the medium correction. Here $\sqrt{z}$ always denotes the root of the complex number $z$ that lies in the upper half plane. 

 From Eq.~(\ref{Pi0})-(\ref{DPi}) we find (from now on we drop the $\perp$ sign from the 2D vectors)
\be
\Pi(\bp,i\omega_\nu) = -i\frac{m_A k_F}{2\pi^2{\alpha_A}^{1/2}\alpha_B^{-3/2}}\int_0^{\pi/2} d\theta \left (\sqrt{i\omega_\nu-\gamma_+(\theta,p)} -\sqrt{i\omega_\nu-\gamma_-(\theta,p) }\right ) 
\label{Pia}
\ee
for $\alpha_A p \leq 1$. Here and in the following
\be
\gamma_\pm (\theta,p) \equiv \alpha_B^{-1}p^2_\pm(\theta)+\alpha_Ap^2-1,
\ee  
where
$
p_\pm(\theta) = \pm \alpha_A p \cos \theta + \sqrt{1-\alpha^2_A p^2 \sin^2 \theta}
$. 
For $\alpha_A p > 1$ we find
\begin{align}
\Pi(\bp,i\omega_\nu) =& -i\frac{m_A k_F}{2\pi^2{\alpha_A}^{1/2}\alpha_B^{-3/2}} \bigg[ \sqrt{i\omega_\nu-(\alpha_Ap^2-1)} +\frac{1}{\pi}\int_0^{\theta_0} d\theta \left (\sqrt{i\omega_\nu-\gamma_+(\theta,p)} -\sqrt{i\omega_\nu-\gamma_-(\theta,p) }\right )  \bigg],
\label{Pib}
\end{align}
where $\theta_0 = \sin^{-1} (1/\alpha_A p)$.

\section{Calculation of  $ \bar \Omega_j$}
\label{NC}
We now determine $ \bar \Omega_j$ given in Eq.~(9) in the main text, which is reproduced below 
\begin{align}
 \bar \Omega_{j} = \frac{1}{\beta}\sum_{m}\int \frac{d^2p}{(2\pi)^2}\mathcal T_{AB}(\bp,i\omega_m)G^\text{A}_{j}(\bp,i\omega_m).
\label{Omegaa}
\end{align}
In terms of the dimensionless momenta and frequencies introduced earlier, we get 
\begin{align}
 \bar \Omega_{j}  & =  2\frac{gm_A}{\beta}\sum_{m}\int \frac{d^2p}{(2\pi)^2}   \frac{1}{\left [1-g\Pi(\bp,i\omega_m)\right ]\left [ i\omega_m-(p^2-1)\right]} \nn \\
 & = \frac{ gm_A}{\pi \beta}\sum_{m}\int_0^\infty dp   \frac{1}{\left [1-g\Pi(\bp,i\omega_m)\right ]\left [ i\omega_m-(p^2-1)\right]}
 \label{Omegajf}
\end{align}
Submitting Eq.~(\ref{Pia}) and (\ref{Pib}) into Eq.~(\ref{Omegajf}), and performing the Matsubara frequency summation, we find for negative 2D-3D scattering length $a_{\rm eff} < 0$ and in the zero temperature limit $\beta\rightarrow \infty$
\begin{align}
  \bar \Omega_j & = 2 a_{\rm eff}\alpha_A^{1/2}\alpha_B^{-1}\mu_A\int_0^1 dp S(p), 
\end{align}
where 
\begin{align}
S(p) &= \frac{1}{1-k_Fa_{\rm eff}\sqrt{\alpha_B}\frac{1}{\pi}\int_0^{\pi/2} d\theta \left [\sqrt{1-p^2 +\gamma_+(\theta,p)} +\sqrt{1-p^2+\gamma_-(\theta,p)} \right ]}.
\label{sp1}
\end{align}

\end{document}